\documentclass[aps,prl,preprint,superscriptaddress,showpacs]{revtex4}

\usepackage{graphicx}
\usepackage{amsmath}
\usepackage{graphicx, subfigure}
\newcommand{\figwidth}{0.47\textwidth}

\begin{document}

\title{Spin nematic state for a spin S=3/2 isotropic non-Heisenberg magnet}

\author{Yu.~A. Fridman}
 \affiliation{V.I. Vernadsky Taurida national university Vernadsky ave., 4,
Simferopol, Ukraine}

\author{O.~A. Kosmachev}
 \affiliation{V.I. Vernadsky Taurida national university Vernadsky ave., 4,
Simferopol, Ukraine}

\author{B.~A. Ivanov}
\email{bivanov@i.com.ua} \affiliation{Institute of Magnetism,
National Academy of Sciences and Ministry of Education of Ukraine,
36(b) Vernadskii avenue, 03142 Kiev, Ukraine }

\date\today

\begin{abstract}
$S=3/2$ system with general isotropic nearest-neighbor exchange
within a mean-field approximation possesses a magnetically ordered
ferromagnetic state and antiferromagnetic state, and two different
spin nematic states, with zero spin expectation values. Both spin
nematic phases display complicated  symmetry break, including
standard rotational break described by the vector-director $\vec
{u}$ and specific symmetry break with respect to the time reversal.
The break of time reversal is determined by non-trivial quantum
averages cubic over the spin components and can be described by unit
``pseudospin'' vector $\vec {\sigma}$. The vectors $\vec {\sigma}$
on different sites are parallel for a nematic state, and $\vec
{\sigma}$'s are antiparallel for different sublattices for an
antinematic phase.
\end{abstract}


\maketitle

Quantum spin systems have provided a wide playground for the quest
of novel types of quantum ordered states and quantum phase
transitions. A number of exotic states has been discovered, the most
well-known examples are the famous Haldane phase in integer spin
antiferromagnetic chains \cite{Haldane}  and quantum spin nematic
states in the spin-1 bilinear-biquadratic isotropic magnets. A
quantum spin nematic state has zero value of dipolar spin order
parameter, $\langle {\vec {S}} \rangle =0$, but spin rotation
symmetry is spontaneously broken due to nontrivial quadrupolar spin
expectation values of the type of $\langle {S_i S_j +S_j S_i }
\rangle $, $i,j=x,y,z$. For $S=1$ spin operators, the quadrupolar
variable have uniaxial symmetry and can be written through the
condition $\langle {(\vec {S},\vec {u})^2} \rangle =0$. The order
parameter can be present via the vector director $\vec {u}$, in
complete analogy to nematic liquid crystals. In the last two
decades, this state has been actively studied for description of
crystalline magnets, see for review \cite{Nagaev,LoktOstr},
including low-dimensional magnets, see recent articles
\cite{IvKolPRB03,ProbAbsNS,SpinNemPRB06}. The interest has got a
considerable impact, motivated by investigation of multicomponent
Bose-Einstein condensates of neutral atoms with nonzero integer
spins \cite{S1nemBEC,BEC_S1,Zhou}. The investigation of spin nematic
states for $S=1$ systems has been performed for two different
physical models, spin lattice system and and Bose gas of interacting
particles with non-zero spin. The results obtained within both
approaches are in a good agreement and complement each other.

A few novel spin nematic states are found for Bose gas of
interacting atoms with spin $S=2$, including the state with
non-uniaxial symmetry \cite{S2}. The question of the possibility of
some kind of spin nematic states for half-integer spins (the minimal
half-integer spin value allowing such state in $S=3/2$) is of large
general importance for physics of Fermi systems. As was found in
Refs.~\onlinecite{S3/2,S3/2prb06,WUPRL05}, high spin fermionic
systems (ultracold Fermi gases) in some parameter region also
exhibit many properties common to that for spin nematic states. On
the best of our knowledge, for higher spins $S>1$, both integer and
half-integer, the only approach based on the direct analysis of gas
of interacting particle has been used.

The aim of this Letter is to develop a mean-field analysis of the
ground state of spin $S=3/2$ isotropic spin Hamiltonian, with
special attention to the spin nematic state, and to discuss the
symmetry and the excitation spectra for such states.

The most general isotropic exchange interaction for pare of spins
$S=3/2$ includes, additionally to Heisenberg bilinear interaction,
non-heisenberg (biquadratic and bicubic) terms as well, which
naturally leads to the model described by the following
nearest-neighbor interaction Hamiltonian:
\begin{equation}
\label{eq1} \hat H=-\sum\limits_{\left\langle {\vec {l},\vec {\delta
}} \right\rangle } {\left[ {J\vec {S}_{\vec {l}} \vec {S}_{\vec
{l}+\vec {\delta }} +K\left( {\vec {S}_{\vec {l}} \vec {S}_{\vec
{l}+\vec {\delta }} } \right)^2+ L\left( {\vec {S}_{\vec {l}} \vec
{S}_{\vec {l}+\vec {\delta }} } \right)^3} \right]}
\end{equation}
where $\vec {S}_{\vec {l}} $ are spin-3/2 operators at the lattice
site $\vec {l}$; $J$, $K$, and $L$ are the exchange constants,
corresponding to the bilinear, the biquadratic, and the bicubic
exchange interactions, respectively, and summation over pares of the
nearest neighbors $\vec {\delta }$ is implied.

The spin-3/2 state $\left| {\psi _l } \right\rangle $ is a
superposition of four basis states  $S_z \left| {\psi _s }
\right\rangle =s \left| {\psi _s } \right\rangle $, with
coefficients $z_{s}$, $s =3/2$, $1/2$, $-1/2$ and $-3/2$. The
essential quantities are the ratios of these complex numbers, and
the system parameter manifold is three-dimensional complex
projective space $\mathrm{CP}^3$. It is convenient to present the
ratios of $z_{s}$ as,
\begin{eqnarray}
\label{eq2} \nonumber \frac{z_{-3/2}}{z_{3/2}} &=&e^{i\varphi }
\cdot \tan \frac{\theta }{2}\,,  \ \frac{z_{-1/2}}{z_{1/2}}
=e^{i\beta }\cdot \tan \frac{\alpha }{2}\,, \\
\frac{z_{1/2}}{z_{3/2}}&=&e^{i\gamma } \cdot \tan \mu \cdot \cos
\frac{\alpha}{2}\cdot \sec \frac{\theta}{2}\, .
\end{eqnarray}

It is easy to check that the resolution of identity $\smallint
D[\psi]\left| \psi \right\rangle \left\langle \psi
\right|=\widehat{1}$, with the proper measure $D[\psi]$, is
satisfied, and this six-parameter state has the properties of SU(4)
coherent states.

Using the states (\ref{eq2}), one can construct the coherent state
path integral and find the effective Lagrangian $\mathcal{L}$
\begin{eqnarray} \nonumber
\label{eq3} &\mathcal{L}& =\hbar \sum_{\vec {l}} \biggl\{\left[
\partial _t \gamma + \left( \partial _t \beta  \right)\sin
^2\frac{\alpha }{2} \right]\sin ^2\mu +
\\ &+& \left( \partial _t \varphi  \right)\sin ^2\frac{\theta
}{2}\cos ^2\mu \biggr\}   - W, \, W = \left\langle \hat H
\right\rangle \, ,
\end{eqnarray}
where $\partial _t= \partial /\partial t$ and $W$ is  the
``classical'' (mean-field) energy of the system, which equals to the
quantum mean value of the Hamiltonian with the states (\ref{eq2}).

The analysis of the ground state can be considerably simplified by
making a rotation in the spin space to the principal axes of the
spin-quadrupolar ellipsoid, i.e. by the use of condition
$\left\langle {S^iS^j+S^jS^i} \right\rangle =0$, if $i\ne j$. This
condition yields the relations $\theta +\alpha =\pi $, $\beta =
-\gamma$, $\varphi =\gamma $, and we can operate with the
three-parameter state
\begin{multline}
\label{3par}
 \left| \psi \right\rangle =\cos \mu \left[ {\sin \frac{\alpha }{2}\left|
{\frac{3}{2}} \right\rangle +e^{i\gamma } \cos \frac{\alpha
}{2}\left|
{-\frac{3}{2}} \right\rangle } \right]+ \\
 +\sin \mu \left[ {\sin \frac{\alpha }{2}\left| {-\frac{1}{2}} \right\rangle
+e^{i\gamma }\cos \frac{\alpha }{2}\left| {\frac{1}{2}}
\right\rangle } \right]\, .
\end{multline}

For this state, diagonal biquadratic expectation values are
independent on $\gamma $ and $\alpha$, they can be written as
\begin{eqnarray}
\label{Squad} \nonumber
 4 \left\langle {S_x^2 } \right\rangle &=&
{3+2\sqrt 3 \sin 2\mu +4\sin ^2\mu } \, , \\
 4\left\langle {S_y^2 } \right\rangle &=&
{3-2\sqrt 3 \sin 2\mu +4\sin ^2\mu } \, .
\end{eqnarray}

The mean values of the spin operators are determined by all three
parameters,
\begin{eqnarray}
\label{eq5} \nonumber \left\langle {S_x + iS_y } \right\rangle
&=&\sin \mu \sin \alpha \cdot \left[ {e^{ - i\gamma }\sin \mu +
e^{i\gamma }\sqrt 3 \cos
\mu } \right]\, ,\\
 2\left\langle {S_z } \right\rangle &=&\cos \alpha \left[ {1-4\cos
^2\mu } \right]\, ,
\end{eqnarray}
and the angular variable $\gamma $ determines the orientation of the
mean value of spin operator $\langle {\vec {S}} \rangle $ in the
$xy-$plane.

Assuming a uniform ground state, one arrives to the following
expression for the energy of the system $W$
\begin{eqnarray}
\label{eq6} \nonumber  W &=& -\biggl( J-\frac{K}{2}+\frac{103}{16}L
\biggr)\biggl[ \frac{\cos ^2\alpha}{4} \bigl( \sin ^2\mu -3\cos
^2\mu \bigr)^2 +
\\&+& \sin ^2\mu \sin ^2\alpha \bigl( 1+2\cos ^2\mu +\sqrt 3 \sin 2\mu
\cos 2\gamma  \bigr) \biggr]\, .
\end{eqnarray}

\begin{figure}[!tb]
\includegraphics[width=\figwidth]{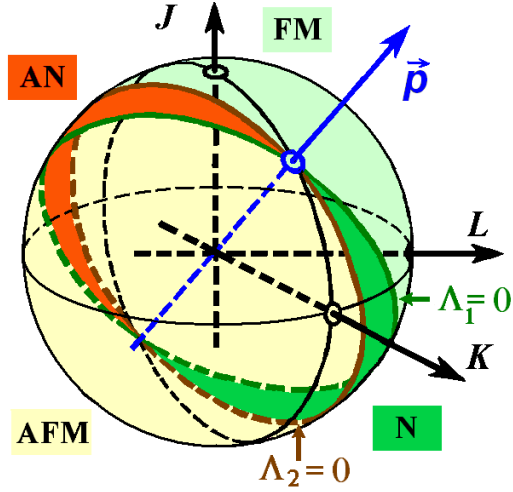}
\caption{ (Color online) The domains of stability for different
phases present through ``angular variables'' (\ref{eq8}).}
\label{fig1}
\end{figure}

In fact, Eq.~(\ref{eq6}) gives us the free energy of the
non-heisenberg $S=3/2$ magnet with the Hamiltonian (\ref{eq1}) at
zero temperature $T$. Minimizing $W$ over parameters $\mu $, $\alpha
$ and $\gamma $, one can find possible states of the system. Our
analysis shows, at $\Lambda _1>0$, where
\begin{equation}
\label{Lam1} \Lambda _1 =16J-8K+103L\, ,
\end{equation}
the minimum corresponds to the ferromagnetic phase with the
saturated spin. For this state $\alpha =\pi $, $\mu =0$, and
\begin{equation}
\label{fm} \left| \Psi  _{\mathrm{ferromagnet}} \right\rangle
=\left| {\frac{3}{2}} \right\rangle , \, \langle {S^z} \rangle
=\frac{3}{2}, \, \langle {S^{x}} \rangle=\langle {S^{y}} \rangle =0.
\end{equation}
Otherwise, at $\Lambda_1<0$ the minimum corresponds to the spin
nematic state with $\alpha =\pi /2$, $ \mu =0$, $\langle {\vec{S}}
\rangle =0$, and
\begin{eqnarray}
\label{ns} \nonumber \left| \Psi _{\mathrm{nematic}} \right\rangle
&=&\frac{1}{\sqrt{2}}\left(\left| {\frac{3}{2}} \right\rangle +
\left| -{\frac{3}{2}} \right\rangle\right),  \\ \langle {(S^z)^2}
\rangle &=& \frac{9}{4},  \, \langle {(S^{x})^2} \rangle = \langle
{(S^{y})^2} \rangle =\frac{3}{4}.
\end{eqnarray}
The image of this state is an elongated ellipsoid of evolution which
can be described by unit vector-director $\vec {u}$, $\vec {u}$ is
parallel to $z-$axis for \eqref{ns}.  So, the quantum phase
transition ``ferromagnetic state -- spin nematic state'' occurs at
$\Lambda _1=0$. In a natural parametrization of the exchange
parameters through ``angular variables'' of the form
\begin{equation}
\label{eq8} J=\tilde {J}\cos \Theta , \ K=\tilde {J}\sin \Theta \cos
\Phi , \ L=\tilde {J}\sin \Theta \sin \Phi \, ,
\end{equation}
the set of system parameters can be presented as a point on a sphere
$J^2+K^2+L^2 =\tilde {J}^2$. The condition $\Lambda_1 = 0$ is
presented by a big circle on the sphere, passing through the
direction $\vec {p}$ with $\tan \Theta =2J/K$ and $\Phi =0$, and the
point with $L=8K/103$ on the equator, see Fig. 1. On this line, all
uniform states of the system are continuously degenerated on the
parameters $\alpha $ and $\mu$. Thus there are lines with high
symmetry in the parameter space of the $S = 3/2$ magnet. Note the
equality $\Lambda _1=0$ corresponds to the condition of lifting of
the symmetry of spin $S=3/2$ Fermi gas in one band Habbard model,
see Eq.(55) of Ref.~\onlinecite{S3/2prb06}.

The complete investigation of the stability of any phase with
respect to arbitrary small perturbation should include an analysis
of evolution of small deviations from the uniform state, $\delta
\alpha_{i, \vec {l}}=\alpha_{i, \vec {l}} - \alpha_{i, 0} $, where
$\alpha_i $ denotes one of six parameters of the state \eqref{eq2},
$\alpha_{i, 0}$ corresponds to the ground state. Following the Bloch
theorem, one can write $\delta \alpha_{\vec {l}}= \Sigma_{\vec
{k}}[\delta \alpha_{\vec {k}}\cdot\exp ( {i\vec {k}\vec {l}} )]$,
$\vec {k}$ takes values within the first Brillouin zone. Information
about stability can be extracted from the spectrum of elementary
excitations $\omega =\omega _i ( {\vec {k}} )$ found for the state
of interest. Such excitations can be learned by use of different
approximate quantum approaches, e.g. bosonization of spin operators
with the $1/N$ expansion \cite{Nikos}, or the Habbard operator
technique, see, for example, \cite{Habbard}. Alternatively, one can
apply a semiclassical treatment based on the Lagrangian (\ref{eq3}),
with the usage of the concrete form of $W$, that is consistent with
mean-field ground state calculations \cite{phenom}. We had used two
last methods for a $d-$dimensional hypercubic lattice, and found (as
for spin $S=1$ magnets, see \cite{phenom}) the same results for the
both approaches.

Variation of the Lagrangian (\ref{eq3}) leads to the Hamilton system
for generalized coordinates $\delta \gamma_{\vec {k}}$, $\delta
\beta_{\vec {k}}$ and $\delta \phi_{\vec {k}}$, its solution gives
three branches of elementary excitations. Here we discuss briefly
the properties of modes important for understanding of the stability
of states.

For the spin nematic state, the properties of two spin wave modes
are common to that for a spin $S=1$ nematic, their spectra are
degenerated with the gapless dispersion law, linear over $k=|\vec
{k}|$ at $k \to 0$, $\omega _{1,2} (\vec {k})\to ck$, where $ \hbar
c= (3a/8)\sqrt {-2z\Lambda_1 \cdot \left( 4K-5L \right) }$. The
third mode also has a gapless spectrum with linear asimptotics at $k
\rightarrow 0$,
\begin{equation}
\label{eq11} \hbar \omega _3 ( {\vec {k}}) = \frac{9}{4}\sqrt{\bar
L_{\vec {k}} \left( -z\Lambda_1 + 16\bar J_{\vec {k}}-8\bar K_{\vec
{k}}+119 \bar L_{\vec {k}}\right)}\,,
\end{equation}
where $z$ is a coordination number, $ \bar J_{\vec {k}}$, $\bar
K_{\vec {k}}$ and $\bar L_{\vec {k}} $ are Fourier-components of
exchange constants of the form
$$\bar J_{\vec
{k}} =J\sum\nolimits_\delta {\left(1-e^{i\vec {k}\vec {\delta }}
\right)} , \bar J_{\vec {k}} \simeq J (a k)^2 \ \mathrm{at}\ k \to 0
\,.  $$
 The phase speed $c_3$ of this mode vanishes at
$L \to 0$, $\hbar c_3 = (9a/4)\sqrt{-zL\Lambda_1}$. For
$\Lambda_1<0$ these modes describe a long-wave instability at the
line (\ref{Lam1}). Additionally, at $\Lambda_2 <0 $, where
\begin{equation}
\label{Lam2} \Lambda _2 =16J-8K+135L\, ,
\end{equation}
the values of $\omega _3$ are imaginary for the quasimomentum $\vec
{k}$ at the edge of Brillouin zone, where $\bar J_k = 2zJ$, ect.
This means the instability, which leads to the transition to
two-sublattice phase (antiferromagnetic, see below) on the line
(\ref{Lam2}). Thus, the spin nematic state is stable if and only
under the conditions $\Lambda_1<0$ and $\Lambda_2>0$.  Note the line
$\Lambda_2 = 0$ describes a big circle, passing through the same
direction $\vec {p}$ as for $\Lambda_1 = 0$, see Fig. 1. Thus, the
nematic state possesses \emph{three Goldstone modes} (not two as for
$S=1$ system). We will discuss this fact below.

The situation for the ferromagnetic state is simpler. One mode
corresponds with the spin oscillations with $| \langle \vec S
\rangle | = 3/2$; it have the gapless dispersion law, parabolic at
$k \to 0$, $16 \hbar \omega _1 ( k )=(ak)^2(48 J+72 K +189 L)$, and
standard for Heisenberg ferromagnets. Two other modes possess
oscillations of $| \langle \vec S \rangle | $. They  have gaps,
proportional to $\Lambda_1$ and describe the long-wave instability
at $\Lambda _1<0$. As well, these modes at $\Lambda_2< 0$ show the
instability with respect to the transition to a two--sublattice
state (an antinematic phase, see below). Thus, ferromagnetic state
is stable if and only $\Lambda_1>0$ and $\Lambda_2>0$.

Let us return to spin  nematic state. An appearance of the third
Goldstone mode \eqref{eq11} for this state can not be explained
within a standard imaging of a nematic state as a ellipsoid of
rotation, connected with bilinear spin expectation values
\eqref{ns}. But for spin  $S=3/2$ magnet, an additional break of
symmetry is caused by non-trivial expectation values \emph{cubic}
over spin components. For the state \eqref{3par}, such values are
$\sigma _{(+)} =(1/3)\langle {\left( {S_x +iS_y } \right)^3} \rangle
$ and $\sigma _{(-)} =(1/3)\langle {\left( {S_x -iS_y } \right)^3}
\rangle $,  $\sigma _{\left( \pm \right)} =\cos ^2\mu \sin \alpha
\exp \left( {\pm i\gamma } \right)$. The transformation properties
of the quantities $\sigma _x=(\sigma _{(+)}+\sigma _{(-)})/2$ and
$\sigma _y=i(\sigma _{(-)}-\sigma _{(+)})/2$ are the same as for the
components of the planar (two-dimensional) vector $\vec {\sigma }$,
$\vec {\sigma } \propto \vec {e}_x \cos \gamma +\vec {e}_y \sin
\gamma $ under the rotation around $z-$axis,  $|\vec {\sigma }|=1$
for the nematic state. Having in mind that $\vec {\sigma }$ changes
its sign at time reversal $t\to -t$, we can say that the properties
of this unit vector are common to that for spin expectation value
and the mode \eqref{eq11} is associated with the oscillations of
this vector.

The presence of such characteristic as $\vec {\sigma }$ is a feature
of principal importance.  As we mentioned above, for a spin $S=1$
nematic, the symmetry breaking is associated with the quadrupolar
ellipsoid only. But the spin $S=3/2$ nematic state is characterized
additionally by unit ``pseudospin'' vector $\vec {\sigma }$, which
changes the sign at time reversal. The additional mode with  $\omega
_3 ( {\vec {k}} )$ has the same properties as for an easy plane
Heisenberg ferromagnet, with the vector $\vec {\sigma }$ playing the
role of the in-plane magnetization. Thus, we arrive to the following
non-trivial picture of the spin-3/2 nematic state: it is
independently SO(3) degenerated over the orientation of the unit
vector-director $\vec {u}$, parallel to the long axis of the
ellipsoid, and SO(2)-degenerated over the direction of unit
pseudospin vector $\vec {\sigma }$, perpendicular to $\vec {u}$. The
presence of this complicated spontaneous symmetry breaking leads to
the appearance of three aforementioned Goldstone modes for this
phase.

This features of uniform nematic state  gives rise for even more
non-trivial properties of multi-sublattice state with $\langle {\vec
{S}} \rangle =0$. For the spin $S=1$, either a two-sublattice
``orthogonal nematic'' with $\left( {\vec {u}_1 \cdot \vec {u}_2 }
\right)=0$ \cite{Nikos}, or a ``threemerized'' states
\cite{threemer}, have been discussed. As we have shown, for the
spin-3/2 system with bipartite lattice, two-sublattice
``antinematic'' state, with parallel $\vec {u}_1 =\vec {u}_2$, but
antiparallel $\vec {\sigma }_1 =-\vec {\sigma }_2 $, is stable in
the region
\begin{equation}
\label{eq12} \Lambda _2 <0,\,\Lambda _1 >0.
\end{equation}

For the rest of the parameter region, $\Lambda _2 <0$, $\Lambda _1
<0$, we have shown the presence of the two-sublattice
antiferromagnetic state with $\langle \vec {S}_1 \rangle =-\langle
\vec {S}_2  \rangle $, $| \langle \vec {S}_{1,2} \rangle  |=3/2$.

The above analysis has been done in the mean-field approximation
only. A rich variety of ``spin-liquid'' states with properties
governed by non-small quantum fluctuations, are known for
one-dimensional (1d) spin-one systems (spin chains), where the
mean-field approximation is not valid. But one can expect for the
spin-3/2 system the role of such fluctuations should be different
from those for a spin-1. In line with Haldane conjecture, for a spin
$S=3/2$ chain one can expect a transformation of the
antiferromagnetic state to a ``critical'' state with gapless
elementary excitations and antiferromagnetic correlations decaying
with power law (compare with gapped Haldane state \cite{Haldane} for
integer spins $S=1,\, 2$ cases). The gapless Luttinger liquid state
for one-dimensional spin $S=3/2$ Fermi gas has been found by Wu
\cite{WUPRL05}. For a spin $S=1$ chain, nematic order is probably
destroyed by non-perturbative quantum fluctuations \cite{ProbAbsNS}.
As has been shown above for the spin $S=3/2$ nematic and antinematic
states the ordering of vector-director $\vec {u}$ is accompanied by
``pseudospin'' $\vec {\sigma }$ ordering. It is clear that the
stability conditions for such states should differ from that for
$S=1$, a time-reversal ``pure nematic'' state. One can expect quite
non-trivial behavior of the nematic and antinematic states for 1d
systems, but a detail investigation of these features is going far
from the scope of this Letter.

Thus, an isotropic magnet with spin $S=3/2$ at $L\neq0$ shows (at
least, in mean-field approximation) the presence of spin nematic and
spin antinematic states with unique dynamic and static properties.
The adequate description of an isotropic magnet with $S=3/2$ needs
consideration of all three possible spin invariants in the exchange
Hamiltonian \eqref{eq1}. Unfortunately, we can not provide a
concrete example of a crystalline magnet with spin $S=3/2$ and high
enough biquadratic exchange. Moreover, to our knowledge, the values
of bicubic exchange constant $L$, important for the presence of
nematic states, has never been discussed for any real compounds. But
the technique of ultracold gases of neutral atoms loaded into
optical lattices gives the possibility of unprecedented control over
the model parameters (exchange integrals, in our consideration).
Alkali atoms $^{132}$Cs, as well as alkaline-earth atoms $^{9}$Be,
$^{135}$Ba, and $^{137}$Ba, having hyperfine 3/2 spin, could be used
for the study of quantum phase transitions described above.

To conclude, as for  well-studied spin $S=1$ models, spin $S=3/2$
system within the mean-field approximation have two magnetically
ordered phases, the ferromagnetic state and the antiferromagnetic
state with maximal possible magnitude of spin on a site. As well,
there are two different nematic states, in which the average spin
equals zero. The domains of the stability for the nematic states
separate the stability regions of the ferromagnetic and the
antiferromagnetic phases, see Fig.~1. On the phase transition lines,
$\Lambda_1=0$ or $\Lambda_2=0$ the symmetry is higher  than the
SO(3)$\sim $SU(2) rotational symmetry of the Hamiltonian. The
qualitative differences between the nematic state for the $S = 3/2$
and  $S =1$ magnets are as follows. The spin $S = 3/2$ nematic
phases display specific symmetry break with respect to the time
reversal. It is exhibited by the properties of the quantum averages
cubic over the spin components, which can be organized in
``pseudospin'' vector $\vec{\sigma}$. By virtue of this fact the
spectrum of spin oscillations in the spin nematic state includes the
third Goldstone (quasiferromagnetic) mode. The antinematic phase has
the antiparallel orientations of $\vec{\sigma}$ at different
sublattices and possesses the following element of symmetry: the
time reflection  combined with the spatial translation on the
nearest neighbor vector $\vec{\delta }$.

\emph{Acknowledgments.--} We are thankful to V.~G. Bar'yakhtar,
A.~K. Kolezhuk and S.~M. Ryabchenko for stimulated discussions. The
work is supported by the grant \# 219 - 09 from Ukrainian Academy of
Science.

\end{document}